\documentclass[10pt,nofootinbib]{revtex4}
\usepackage{amssymb}
\usepackage{latexsym}
\usepackage{epsfig}
\usepackage{float,rotating}
\usepackage{multirow}
\usepackage{mathtools}\usepackage{bbm}
\usepackage{graphicx}
\usepackage[utf8]{inputenc,xcolor}
\usepackage{graphicx}
\usepackage{array}
\usepackage{multirow,bigdelim}
\usepackage{filecontents}

\usepackage{booktabs,cancel,bigints}
\usepackage{subfigure}
\begin{document}
\title{Topological dyonic dilaton black holes in AdS spaces}
\author{S. Hajkhalili$^{1}$\footnote{hajkhalili@gmail.com} and A. Sheykhi$^{1,2}$\footnote{asheykhi@shirazu.ac.ir}}
\address{$^1$ Physics Department and Biruni Observatory, College of Sciences, Shiraz University, Shiraz 71454, Iran\\
$^2$ Research Institute for Astronomy and Astrophysics of Maragha
(RIAAM), P.O. Box 55134-441, Maragha, Iran}
\begin{abstract}
We construct a new class of dyonic dilaton black hole solutions in
the background of Anti-de Sitter (AdS) spacetime. In order to find
an analytical solution which satisfy all the field equations, we
should consider the string case where the dilaton coupling is
$\alpha=1$. The asymptotic behaviour of the solution
($r\rightarrow\infty$) is exactly AdS, where the dilaton field
becomes zero and the metric function reduces to $f(r)\rightarrow
-\Lambda r^2/3$. In this spacetime, black hole horizons and
cosmological horizons, can be a two-dimensional positive, zero or
negative constant curvature surface. We study the physical
properties of the solution and show that depending on the metric
parameters, these solutions can describe black holes with one or
two horizons or a naked singularity. We investigate thermodynamics
of the solutions by calculating the charge, mass, temperature,
entropy and the electric potential of these solutions and disclose
that these conserved and thermodynamic quantities satisfy the
first law of black hole thermodynamics. We also analyze thermal
stability of the solutions and find the conditions for which we
have a stable dyonic dilaton black hole.


\end{abstract}
\maketitle
\section{Introduction}
The first attempt to consider a minimal coupling between the
scalar fields and gravity was done by Fisher, who found a static
and spherically symmetric solution of the Einstein-massless scalar
field equations \cite{fisher}. A great varieties of the scalar
fields have been examined for the different goals. One of the
interesting scalar field in the black hole physics is the dilaton
field. This scalar field appears in variety theories such as
dimensional reductions, low-energy limit of string theory, various
models of supergravity. A dilaton field naturally couples to gauge
fields and thus can influence various physical phenomena. For
instance, the presence of the dilaton field can change the causal
structure and asymptotic behavior of the spacetime as well as the
stability and thermodynamical features of the charged black holes.
Given the asymptotic behavior of black holes, various theories can
be defined for understanding their physics as well as the Universe
behavior. For instance, a powerful tool to get insight into the
strong coupling dynamics of certain field theories by studying
classical gravity is AdS/CFT which is a correspondence between the
gravity in an AdS spacetime and conformal field theory (CFT)
living on the boundary of the AdS spacetime \cite{ads}. The
asymptotically flat dilaton black hole solutions in which the
dilaton field is coupled to the Maxwell field in an exponential
way have been explored in \cite{string1}. It was shown that a
asymptotically nonflat, non (A)dS solutions can be constructed
using one or two Liouville-type dilaton potential \cite{string2}.
In order to obtain an asymptotically (A)dS solution in the
presence of dilaton field, the combination of three Liouville type
dilaton potentials were required \cite{string3,string4}. In this
case, the potential includes three exponential dilaton field
$(\Phi)$ such as \cite{gao1,gao2}
\begin{equation}\label{Vphi}
V(\Phi)={\frac {2\Lambda}{ 3\left( {\alpha}^{2}+1 \right) ^{2}}
\left[8\,{\alpha}^{2}{{\rm
e}^{\Phi\left(\alpha\,-{{1}/{\alpha}}\right)}} + {\alpha}^{2}
\left( 3\,{\alpha}^{2}-1 \right) {{\rm e}^{-\,{
{2\Phi}/{\alpha}}}} - \left( {\alpha}^{2}-3 \right) {{\rm
e}^{2\,\alpha\, \Phi}}\right] },
\end{equation}
where $\alpha$ is a constant which determines the strength of the
coupling between dilaton field and gauge field. A special case
occurs when the coupling constant is $\alpha=1$ that appears in
the low energy limit of string action in Einstein's frame and
corresponds to the bosonic sector of the field theory limit of the
superstring model ($N=4$ in supergravity) which calls the "string"
conformal gauge \cite{string4}. Also the authors of \cite{string5}
showed that one of the best case in dilaton gravity to obtain the
generalized Toda lattices related to Lie algebra is the string
case where $\alpha=1$. The properties of electrically charged
dilaton black holes in the background of AdS spacetime have been
investigated \cite{AdSD}. If a black hole contains both electric
and magnetic charges it is called dyonic black hole. This type of
black holes are one of the best objects to study the effect of
external magnetic field on superconductors as well as Hall
conductance and DC longitudinal conductivity \cite{dyonic1}. Also,
it has been argued that dyonic black hole in AdS background may be
holographic dual of van der Waals fluid with chemical potential
\cite{dyonic2} and their dual on the conformal boundary of AdS
spacetime is the stationary solutions of relativistic
magnetohydrodynamics equations \cite{dyonic3}.

In the present work, we would like to find asymptotically AdS
dyonic black hole solutions in the presence of dilaton field. Many
papers are devoted to investigate dilaton dyonic black holes
\cite{dyonic4}. However, their asymptotic behaviour are neither
AdS nor dS. Also, in \cite{dyonic5} dyonic black hole in the
Kaluza-Klein theory in the presence an additional scalar potential
was explored. In the absence of dilaton field, dyonic black holes
in the background of AdS spacetime have been investigated in
\cite{dyonic6,drhendi}. The solution of massive gravity in the
presence of electric and magnetic charges was explored in
\cite{dyonic7}.

This paper is organized as follows. In section \ref{Field}, we
introduce the action and find the basic field equations by varying
the action. In section \ref{temP}, we study thermodynamical
properties of the solutions. In section \ref{stab}, thermal
stability of the solutions are checked. We finish our paper with
closing remarks in section \ref{conclus}.
\section{Action and Lagrangian}\label{Field}
\begin{figure}[H]
\centering \subfigure[$p=0.5$, $q=0.61$ and
$b=0.2$]{\includegraphics[scale=0.35]{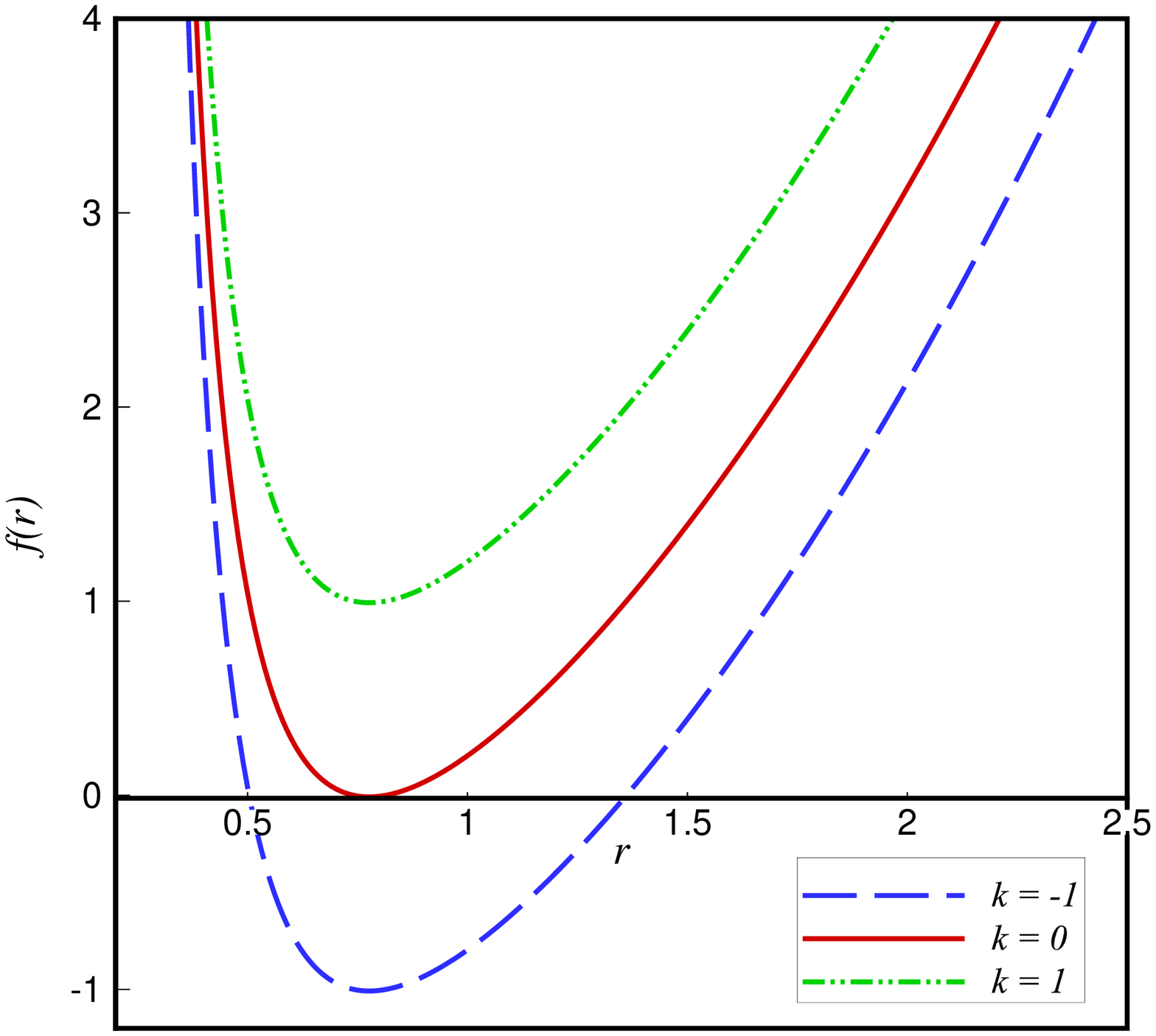}\label{fig1a}}
\hspace*{.05cm} \subfigure[$k=1$, $q=0.7$ and $b=0.2$
]{\includegraphics[scale=0.35]{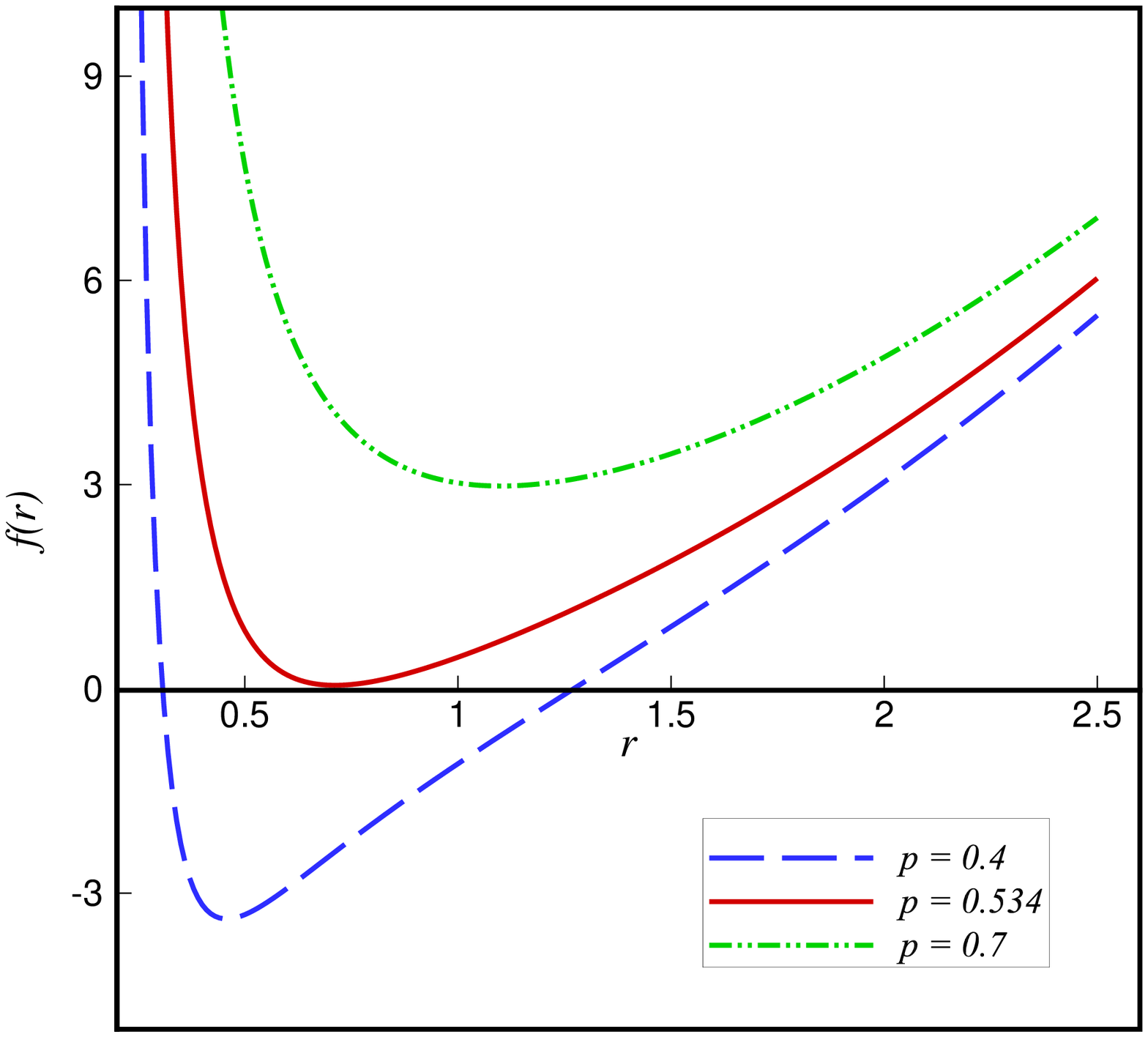}\label{fig1b}} \hspace*{.05cm}
\subfigure[$k=1$, $p=0.5$ and $b=0.2$
]{\includegraphics[scale=0.35]{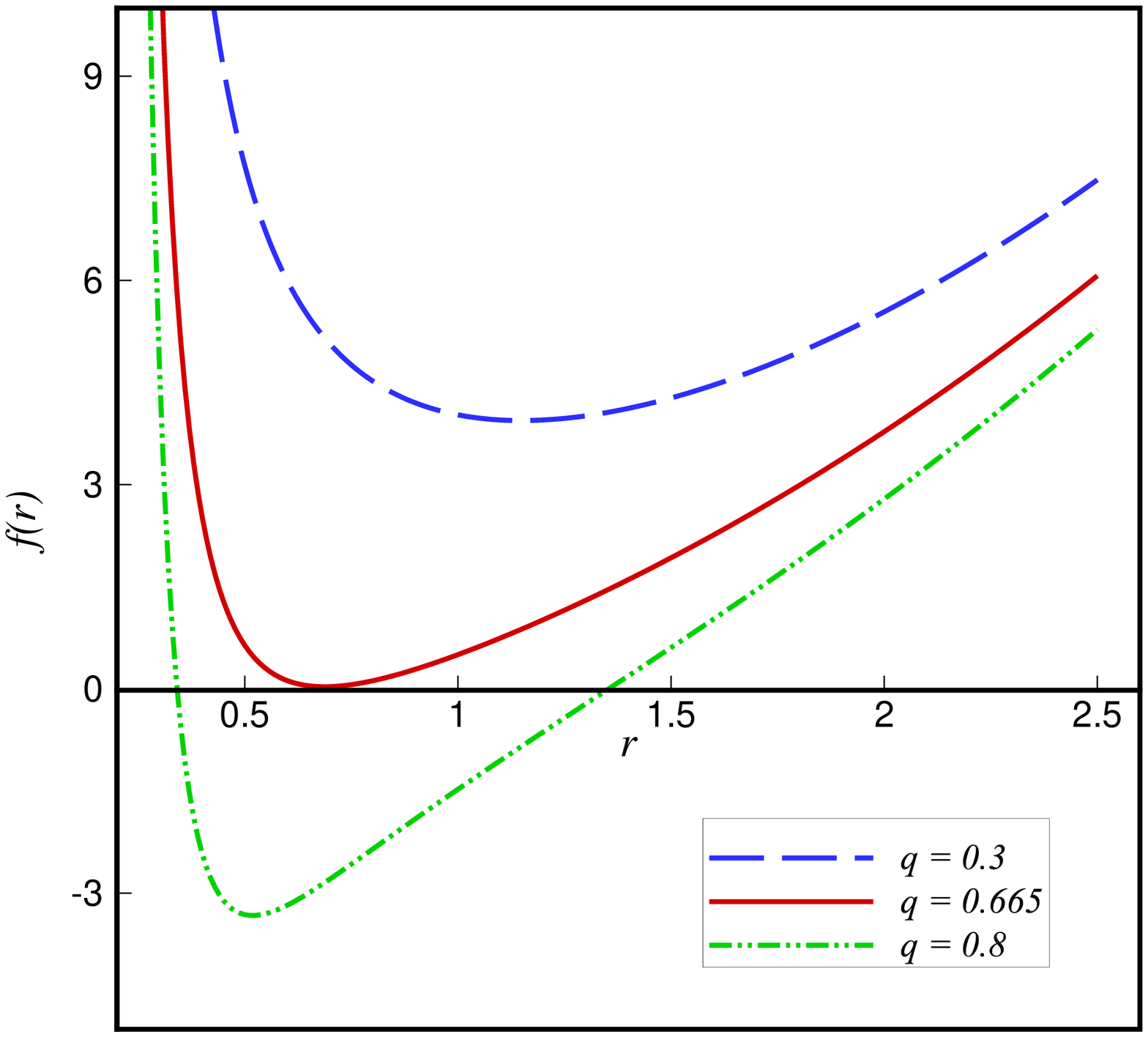}\label{fig1c}} \hspace*{.05cm}
\subfigure[$k=1$, $p=0.4$ and $q=0.56$
]{\includegraphics[scale=0.35]{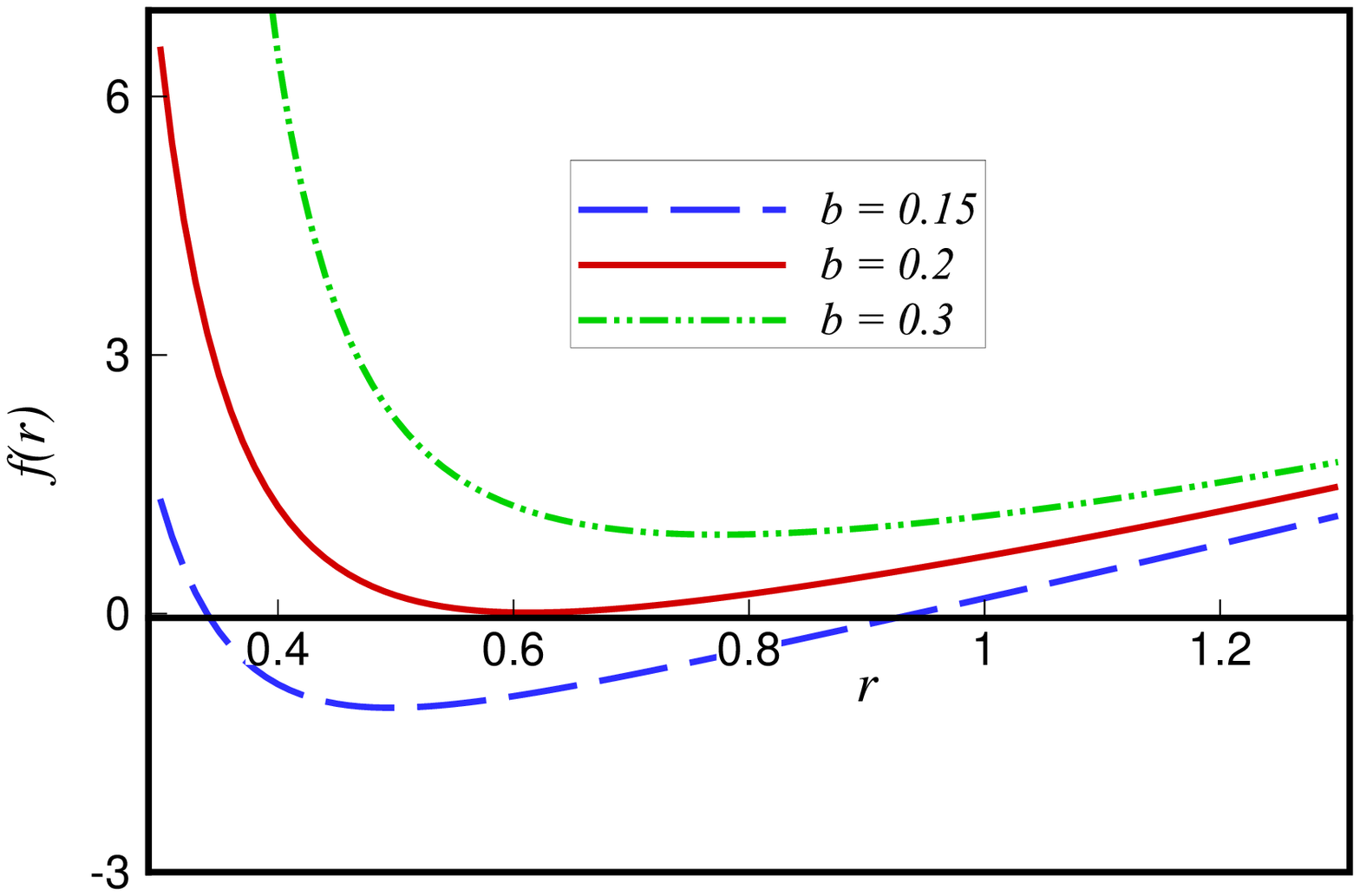}\label{fig1d}} \caption{The
behavior of the metric function $f(r)$ versus $r$.}\label{fig1}
\end{figure}

The action of Einstein-Maxwell-dilaton gravity in the presence of
a dilaton potential is given by
\begin{equation}\label{Act}
S=\frac{1}{16\pi}\int{d^{4}x\sqrt{-g}\Big(\mathcal{R}\text{ }-2
g^{\mu\nu}
\partial_{\mu} \Phi \partial_{\nu}\Phi -V(\Phi
)-e^{-2\Phi}F^2\Big)},
\end{equation}
where Ricci scalar curvature is displayed by $\mathcal{R}$ and
$F^2=F_{\mu\nu}F^{\mu\nu}$ is the usual Maxwell Lagrangian. Also,
$V(\Phi)$ is the potential for $\Phi$ which is given by Eq.
(\ref{Vphi}) by considering the string case ($\alpha=1$),
 \begin{equation}\label{pot}
 V(\Phi)=\frac{\Lambda}{3}\left(4+e^{2\Phi}+e^{-2\Phi}\right)=\frac{2}{3}\Lambda\left(2+\cosh(2\Phi)\right),
\end{equation}
where $\Lambda $ is the cosmological constant. The reason for
considering the string case ($\alpha=1$) comes from the fact that
we could only find an analytical solution in this case and it is
impossible to find an analytical solution for an arbitrary
$\alpha$ which satisfy all the field equations. One may ignore the
dilaton filed, so the potential reduced to $V(\Phi=0)=2\Lambda$
which is the famous term in the action of AdS spaces. Variation of
the action (\ref{Act}) respect to the metric, Maxwell and dilaton
fields, respectively, eventuate
\begin{equation}\label{eqmetr}
{\cal R}_{\mu\nu}=2 \partial _{\mu }\Phi
\partial _{\nu }\Phi +\frac{1}{2}g_{\mu \nu
}V(\Phi)+2e^{-2\Phi}\Big(F_{\mu\beta}F_\nu^\beta-\frac{1}{4}g_{\mu\nu}F^2\Big),
\end{equation}
\begin{equation} \label{eqmax}
\partial_\mu\Big(\sqrt{-g}e^{-2\Phi }F^{\mu\nu}\Big)=0,
\end{equation}
\begin{equation} \label{eqdilaton}
\partial_\mu\partial^{\mu}\Phi=\frac{1}{4}\frac{\partial V}{\partial \Phi}-\frac{1}{2}e^{-2\Phi}F^2.
\end{equation}
Since we are going to construct topological static and spherically
symmetric black holes in four dimensions, we take the following
form for the metric
\begin{equation}\label{metric}
ds^2=-f(r)dt^2 +{dr^2\over f(r)}+ r^2R^2(r)d\Omega_{k}^2 ,
\end{equation}
where $f(r)$ and $R(r)$ are functions of $r$ which should be
determined and $d\Omega_{k}^2$ is the metric of a $2$-dimensional
hypersurface with constant curvature. Hereafter, we take $k =\pm1,
0$, without loss of generality. The explicit form of
$d\Omega_{k}^2$  is
\begin{equation}\label{met}
d\Omega_k^2=\left\{
\begin{array}{ll}
$$d\theta^2+\sin^2\theta d\phi^2$$,\quad \quad\!\!{\rm for}\quad $$k=1$$, &  \\
$$d\theta^2+\theta^2 d\phi^2$$,\quad\quad\quad {\rm for}\quad $$k=0$$,&  \\
$$d\theta^2+\sinh^2\theta d\phi^2$$, \quad {\rm for}\quad $$k=-1$$.&
\end{array}
\right.
\end{equation}
In order to construct dyonic dilaton black hole, we first find the
electric and magnetic fields using the Maxwell equation
(\ref{eqmax}). The only nonzero components of the electromagnetic
field tensor are
\begin{equation}\label{ftr}
F_{tr}=-F_{rt}=\frac{qe^{2\Phi(r)}}{r^2R^2(r)},
\end{equation}
\begin{equation}\label{fthp}
F_{\theta\phi}=-F_{\phi\theta}=\left\{
\begin{array}{ll}
$$p\sin\theta$$,\quad\quad\!\!{\rm for}\quad $$k=1$$, &  \\
$$p$$,\quad\quad\quad\quad\, {\rm for}\quad $$k=0$$,&  \\
$$p\sinh\theta$$, \quad {\rm for}\quad $$k=-1$$.&
\end{array}
\right.
\end{equation}
Here, we denote the electric and magnetic charge by $q$ and $p$,
respectively. Using Maxwell electromagnetic tensor, one can find
that $F_{tr}$ is the electric field and $F_{\theta\phi}$ is the
magnetic field. It is worth noting that the topological structure
of the event horizon does not affect the electric field
(\ref{ftr}), while it affects the functional form of the magnetic
field (\ref{fthp}). Also, respect to the values of the $r$ the
electric field can change, but magnetic field is independent of
$r$ in all three topology. In order to find the metric functions,
we try to solve two remaining field equations (\ref{eqmetr}) and
(\ref{eqdilaton}) by considering the metric function
(\ref{metric}) and the non-zero components of the electromagnetic
tensor (\ref{ftr}) and (\ref{fthp}).  We find the ($tt$) and
($rr$) components of Eq. (\ref{eqmetr}) as follows:
\begin{eqnarray}
&&Eq_{tt}=\frac{f^{\prime\prime}}{2}+\left(\frac{R^\prime}{R}+\frac{1}{r}\right)f^\prime-
\frac{q^2e^{2\Phi}+p^2e^{-2\Phi}}{r^4R^4}+\frac{V(\Phi)}{2}=0,\label{tt}\\[10pt]
&&
Eq_{rr}=\frac{f^{\prime\prime}}{2}+\left(\frac{R^\prime}{R}+\frac{1}{r}\right)
f^\prime+2\left(\Phi^\prime+\frac{r\,R^{\prime\prime}+2R^\prime}{r\,R}\right)f-\frac{q^2e^{2\Phi}+p^2e^{-2\Phi}}{r^4R^4}+\frac{V(\Phi)}{2}=0,\label{rr}
\end{eqnarray}
where the prime denotes derivative with respect to $r$. It is a
matter of calculations to show that the subtraction of Eqs.
(\ref{tt}) and (\ref{rr}) leads to the following equation,
\begin{eqnarray}\label{eqphi}
R^{\prime\prime}+\frac{2}{r}R^\prime+{\Phi^{\prime}}^2~R=0.
 \end{eqnarray}
Now, we make the ansatz
\begin{equation}\label{ansatz}
R(r)=e^{\Phi(r)}.
\end{equation}
Using this ansatz, one can solve Eq. (\ref{eqphi}), which admits a
solution of the form
\begin{equation}\label{phi}
\Phi(r)=\frac{1}{2}\ln\left(1-\frac{b}{r}\right),
\end{equation}
where $b$ is the integration constant which is related to the
dilaton field so that if we set $b=0$, the dilaton field will be
disappeared. In the asymptotic region where $r\rightarrow\infty$,
the dilaton field goes to zero. This is an expected result, since
our solutions are asymptotically AdS and thus the effects of the
dilaton field in the asymptotic region should be removed. Besides,
from (\ref{phi}) we see that the solutions only exist for $r>b$.
Thus, one may exclude the region $r<0 b$ from the spacetime by
redefinition of the coordinate $r\rightarrow \rho$ such that
$\rho=\sqrt{r^2-b^2}$. Thus in the new coordinate $\rho=0$ is
correspondence to $r=b$, the spacetime is regular for $\rho>0$,
and the line elements of the metric may be written as
\begin{equation}
ds^2=-f(\rho)dt^2+\frac{\rho^2\,d\rho^2}{(\rho^2+b^2)f(\rho)}+(\rho^2+b^2)R^2(\rho)(d\theta^2+\sin^2\theta\,
d\phi^2).
\end{equation}
In order to find the metric function $f(r)$, we use the
($\phi\phi$) component of Eq. (\ref{eqmetr}) which is written as
\begin{equation}\label{pp}
Eq_{\phi\phi}=\left(\frac{R^\prime}{R}+\frac{1}{r}\right)f^\prime+
\left(\frac{R^{\prime\prime}}{R}+\frac{{R^\prime}^2}{R^2}+\frac{4R^\prime}{r\,R}
+\frac{1}{r^2}\right)f+\frac{q^2e^{2\Phi}+p^2e^{-2\Phi}}{r^4R^4}+\frac{V(\Phi)}{2}-\frac{k}{r^2\,R^2}=0.
\end{equation}
Substituting potential (\ref{pot}), the ansatz (\ref{ansatz}) and
the dilaton field (\ref{phi}) in the above equation, one can find
\begin{equation}\label{f(r)}
f(r)=-\frac{\Lambda r^2}{3}\, \left( 1-\frac{b}{r}
\right)+\,{\frac {{2p }^{2}}{ \left( b-2\,r \right)  \left( b-r
\right) }}-\,{\frac {{2q}^{ 2}}{ \left( b-2\,r \right) r}}+{\frac
{2(m-k~r)}{b-2\,r}},
\end{equation}
where $m$ is the integration constant which is related to the mass
of black hole. One can also rewrite the above solution in term of
the new coordinate $\rho$ by replacing $r=\sqrt{\rho^2+b^2}$.  In
the asymptotic region ($r\rightarrow\infty$), the dominant term in
(\ref{f(r)}) is the cosmological term,
$f(r\rightarrow\infty)=-\Lambda r^2/3$ which confirms that our
solutions are asymptotically AdS. In the absence of the magnetic
charge $(p=0)$, our solution reduces to the topological dilaton
black hole in AdS spaces for $\alpha=1$
\cite{gao1,SheAdS2,SheAdS3}. Also, in the absence of the dilaton
field ($b=0$), the metric (\ref{f(r)}) reduces to the dyonic black
hole in the AdS spacetime \cite{Dutta,drhendi},
\begin{equation}
f(r)\Big|_{b=0}=k-{\frac {m}{r}}-\frac{\Lambda r^2}{3}+{\frac
{{q}^{2}}{{r}^{2}}}+{\frac {{p}^{2}}{{r}^{2}}}.
\end{equation}
The obtained solutions will fully satisfy all components of the
field equations provided,
\begin{equation}\label{cond}
m=\frac{b}{2}+\frac{2}{b}\left(q^2-p^2\right).
\end{equation}
This condition indicates that the electric and magnetic charges
$(q,p)$ are not independent parameters. Inserting condition
(\ref{cond}) in the metric function (\ref{f(r)}), we find
\begin{equation}\label{f1}
f(r)=k-\frac{\Lambda r^2}{3}\, \left( 1-\frac{b}{r} \right)-{\frac
{2{q}^{2}}{br}}+{\frac {2{p}^{2}}{b \left(r -b\right) }}.
\end{equation}
Next we look for the singularity and horizon(s) of the spacetime.
The divergency in the Ricci and Kretschmann scalars confirm the
existence of the essential singularity. The Ricci scalar is
calculated as
\begin{equation}\label{Ricci}
Ricci=-{\frac { \left( {b}^{2}-8 br+8{r}^{2} \right) \Lambda}{ 2
\left( b-r \right) r}}-{\frac {b{q}^{2}}{{r}^{3} \left( b-r
\right) ^
 {2}}}-{\frac {{p}^{2}b}{{r}^{2} \left( b-r \right) ^{3}}}+{\frac
{k{b}^{2}}{2{r}^{2} \left( b-r \right) ^{2}}}.
\end{equation}
As one can see from Eq. (\ref{Ricci}), the surface $r = b$
$(\rho=0)$ as well as $r=0$ are curvature singularities. Since we
have excluded $r<b$ from our spacetime, thus, in our new
coordinates, the essential singularity is located at $\rho=0$
($r=b$). As we already mentioned our solution only exist for
$r>b$. The explicit form of the Kretchmann invariant is very
complicated and for the economic reasons we do not present it
here. It is easy to show that
\begin{equation}\label{Kret}
\lim\limits_{\rho\rightarrow
0}R_{\mu\nu\alpha\beta}R^{\mu\nu\alpha\beta}\rightarrow \infty.
\end{equation}
From Eqs. (\ref{Ricci}) and (\ref{Kret}), we conclude that there
is an essential singularity at $\rho=0$. The behaviour of these
scalars in the large $\rho$ limit is interesting, too. At this
limit, the dominate term in the Ricci and Kretchmann scalars are
given by
\begin{eqnarray}
\lim\limits_{\rho\rightarrow \infty}R_{\mu\nu\alpha\beta}R^{\mu\nu\alpha\beta}&\rightarrow&\frac{8\Lambda^2}{3},\nonumber\\[10pt]
\lim\limits_{\rho\rightarrow \infty}\rm
Ricci&\rightarrow&4\Lambda,
\end{eqnarray}
which are the characterization of the (A)dS spaces. Now we look
for the horizon(s). Black hole horizon(s) is indeed the positive
real root of Eq. $f(r)=0$. Unfortunately, it is not easy to find
the location of the horizons, analytically. However, we can find
the location of horizons by plotting the function $f(r)$ versus
$r$. In all figures, we set $\Lambda=-3$. According to Fig.
\ref{fig1} the number of horizons depend on the values of the
metric parameters. Our solution may have one or two horizons and
in some cases, it may describe a naked singularity depending on
the parameters. In summary, our solutions can describe black holes
in the case $b< r_{+}$, where $r_{+}$ is the outer horizon of the
black string (the largest root of Eq. $ f(r)=0$).

In Fig. \ref{fig1a} we consider the effect of the horizon topology
on the number of horizons. By fixing other metric parameters, the
number of horizons decreases by increasing $k$. Figs. \ref{fig1b}
and \ref{fig1c} show that with increasing the value of the
magnetic charge, the number of horizons decrease, while  by
increasing the electric charge the number of horizons can
increase, too.
\section{Thermodynamical properties}\label{temP}
The Hawking temperature is attributed to the black hole horizon
(surface gravity $\kappa$) and is defined as
\begin{equation}
T=\frac{\kappa}{2\pi}=\frac{1}{2\pi}\sqrt{-\frac{1}{2}(\nabla_\mu\chi_\nu)(\nabla^\mu\chi^\nu)},
\end{equation}
where $\chi=\partial/\partial t$ is null Killing vector of the
horizon. It is a matter of calculation to show
\begin{equation}\label{temp}
T=\frac{f^\prime(r)_|{_{r=r_+}}}{4\pi}={\frac {1}{2\pi } \left[ {\frac {k}{2r_{_+}}}-
{\frac {{p}^{2}}{r^3}\left( 1-\frac{b}{r_{_+}} \right) ^{-2}}+\frac{\left( 2\,b-3\,r_{_+} \right)}{6} \Lambda
 \right] }.
\end{equation}
Here $$r_{+}=\sqrt{\rho_{+}^2+b^2}$$ shows the outer horizon which
is the largest positive real root of $f(r_{+})=0$. It is also
interesting to note that in the absence of magnetic charge
($p=0$), it reduces to the temperature of charged AdS dilaton
black hole \cite{SheAdS3} for $\alpha=1$. The entropy of the
dilaton black hole typically obeys the area law of the entropy
which is a quarter of the event horizon area \cite{Beckenstein}.
For our solution it is obtained as
\begin{equation}
S=\frac{\Omega_{k}~r_{_+}^2}{4}\left(1-\frac{b}{r_{_+}}\right),
\end{equation}
where $\Omega_k$ is the volume of the two dimensional hypersurface
with constant curvature. The above expression is exactly the
entropy of charged AdS dilaton black hole \cite{SheAdS3}. Using
Brown and York formalism we calculate the mass of the
asymptotically AdS dyonic dilaton black hole \cite{SheAdS3}. We
find
\begin{equation}
M=\frac{\Omega_{k}~(q^2-p^2)}{4~b\pi}=\frac{\Omega_{k}}{8\pi}\left(m-\frac{b}{2}\right).
\end{equation}
In the absence of magnetic charge ($p=0$), it recovers the mass of
the  AdS dilaton black hole \cite{SheAdS3}, while in the absence
of dilaton field ($b=0$), it reduces to the mass of topological
dyonic AdS black holes. One may use the Gauss's law to calculate
the total electric and magnetic charge of the black hole.
According to the Gauss theorem, the electric charge of the black
hole is
\begin{equation}
Q=\frac{1}{4\pi}\int_{r\rightarrow\infty}\sqrt{-g}F_{tr}d^2x=\frac{
q  \Omega_k}{4\pi}.
\end{equation}
Similarly, we can obtain the total magnetic charge of the dyonic
black hole as
\begin{equation}
P=\frac{p \Omega_k}{4\pi}.
\end{equation}
Also, one can obtain $U_{_Q}$ and $U_{_P}$ which are,
respectively, the electric and magnetic potential by applying
their definition using the free energy, which is given as
\cite{Dutta}
\begin{equation}
W=\frac{I_{\rm on shell}}{\beta}
\end{equation}
where $I_{\rm on shell}$ is the on shell action and $\beta$ is the
inverse of temperature. Multiplying both sides of Eq.
(\ref{eqmetr}) by $g^{\mu\nu}$, we arrive at
\begin{equation}\label{R}
\mathcal{R}=2\partial^\mu\phi\partial_{\mu }\phi+2V(\Phi).
\end{equation}
Substituting Eq. (\ref{R}) in Eq. (\ref{Act}), we find
\begin{equation}\label{onshell}
I_{\rm on shell}=\frac{1}{16\pi}\int{d^{4}x\sqrt{-g}\Big(V(\Phi
    )-e^{-2\Phi}F^2\Big)}.
\end{equation}
As $r$ goes to infinity, the above integration diverges. In order
to remove the divergency, one may add some counterterm to the
original action or subtracting the contribution of a background
spacetime. Finally, we drive the electric and magnetic potential
as
\begin{eqnarray}
&&U_{_Q}=-\frac{\partial W}{\partial Q}=\frac{Q}{r},\label{electric pot}\\&&
U_{_P}=\frac{\partial W}{\partial P}=\frac{P}{r-b}.\label{magnetic pot}
\end{eqnarray}
It is important to note that the dialton field does not affect the
electric potential, while it changes the magnetic potential. In
the absence of the dilaton field ($b=0$), magnetic potential is
the same as that in \cite{Dutta,drhendi}. In the thermodynamics
consideration, the satisfaction of the first law of thermodynamics
implies the correctness of conserved and thermodynamic quantities.
In order to check this, we obtain the mass $M$ as a function of
extensive quantities $S$, $Q$ and $P$. We find
\begin{equation}
M(S,Q,P)=\frac{4\pi}{b}(Q^2-P^2).
\end{equation}
Considering the fact that all thermodynamic quantities should be
defined on the horizon, we write $b=b(Q,r,P)$. Taking into account
the fact that the entropy is a function of $S=S(r,b)$ and the mass
is a function of extensive parameters $M=M(S,Q,P)$, we define the
temperature, electric and magnetic potentials as
 \begin{eqnarray}
T=\left(\frac{\partial M}{\partial
S}\right)_{Q,P}=\frac{\left(\frac{\partial M}{\partial
b}\right)_{Q,P}\times dbr}{\left(\frac{\partial S}{\partial
r_{_+}}\right)_{Q,P}+\left(\frac{\partial S}{\partial
b}\right)_{Q,P}\times dbr},
\end{eqnarray}
\begin{equation}
 U_{_Q}=\left(\frac{\partial M}{\partial Q}\right)_{S,P}+\left(\frac{\partial M}{\partial b}\right)_{S,P}\left(\frac{\partial b}{\partial
 Q}\right)_{S,P},
\end{equation}
\begin{equation}
U_{_P}=\left(\frac{\partial M}{\partial
P}\right)_{S,Q}+\left(\frac{\partial M}{\partial
b}\right)_{S,Q}\left(\frac{\partial b}{\partial P}\right)_{S,Q},
\end{equation}
where we can compute $dbr$ as
\begin{equation}
dbr=\left(\frac{db}{dr_{_+}}\right)_{Q,P}=-\frac{\left(\frac{\partial f}{\partial r}\right)_{Q,P}}{\left(\frac{\partial f}{\partial b}\right)_{Q,P}
}=\frac{{{\frac {2{q}^{2}}{b{r}^{2}}}-{\frac {2{P}^{2}}{b \left( -r+b \right) ^{2}}}-\frac{\Lambda \left( -2r+b \right) }{3}
 }}{{\frac {2{q}^{2}}{{b}^{2}r}}+{\frac {2P^2(-r+2\,b)}{{b}^{2} \left( -r+b \right) ^{2}}}+\frac{\Lambda ~r}{3}
 }.
\end{equation}
Finally, our calculations show that all intensive quantities given
in Eqs. (\ref{temp}), (\ref{electric pot}) and (\ref{magnetic
pot}) satisfy the first law of black hole thermodynamics,
\begin{equation}
dM=TdS+U_{_Q} dQ+U_{_P}dP.
\end{equation}
\section{Thermal Stability}\label{stab}
This section is devoted to study the stability of the dyonic
dilaton black holes. Since we consider three extensive variables
$X_i$, the best way to check the stable range of our solution is
to work in the grand-
canonical ensemble and study the Hessian
matrix \cite{hess} which is defined as
\begin{equation}\label{hessMat}
H^{^M}_{_{X_i X_j}}=\frac{\partial^2M}{\partial X_i\partial X_j}.
\end{equation}
Positivity of determinant of the Hessian matrix is sufficient to
ensure thermal stability in the grand-canonical ensemble. As
mentioned before we may select $X_i=(S,Q,P)$. Since the mass is
not an explicit function of $X_i$, we use the following
definitions to obtain the components of the Hessian matrix,
\begin{eqnarray}
&&H_{11}=\left(\frac{d^2M}{d^2S}\right)_{P,Q}=\left(\frac{dT}{dS}\right)_{P,Q}=\frac{\left(\frac{\partial T}{\partial r_{_+}}\right)_{P,Q}+\left(\frac{\partial T}{\partial b}\right)_{P,Q}\times dbr}{\left(\frac{\partial S}{\partial r_{_+}}\right)_{P,Q}+\left(\frac{\partial S}{\partial b}\right)_{P,Q}\times dbr},\nonumber\\[10pt]&&
H_{22}=\left(\frac{d^2M}{d^2P}\right)_{S,Q}=\left(\frac{dM_P}{dP}\right)_{S,Q}=\left(\frac{\partial M_P}{\partial P}\right)_{S,Q}+\left[\left(\frac{\partial M_P}{\partial r}\right)_{S,Q}\left(\frac{dr_{_+}}{db}\right)_{S,Q}+\left(\frac{\partial M_P}{\partial b}\right)_{S,Q}\right]\left(\frac{dP}{db}\right)^{-1}_{S,Q},\nonumber\\[10pt]&&
H_{33}=\left(\frac{d^2M}{d^2Q}\right)_{S,P}=\left(\frac{dM_Q}{dQ}\right)_{S,P}=\left(\frac{\partial M_Q}{\partial Q}\right)_{S,P}+\left[\left(\frac{\partial M_Q}{\partial r}\right)_{S,P}\left(\frac{dr_{_+}}{db}\right)_{S,Q}+\left(\frac{\partial M_Q}{\partial b}\right)_{S,P}\right]\left(\frac{dQ}{db}\right)^{-1}_{S,P},\nonumber\\[10pt]
&&H_{12}=H_{21}=\left(\frac{d^2M}{dSdP}\right)_{Q}=\left(\frac{dT}{dP}\right)_{S,Q}=\left(\frac{\partial T}{\partial P}\right)_{S,Q}+\left[\left(\frac{\partial T}{\partial r_{_+}}\right)_{S,Q}\left(\frac{dr_{_+}}{db}\right)_{S,Q}+\left(\frac{\partial T}{\partial b}\right)_{S,Q}\right]\left(\frac{dP}{db}\right)^{-1}_{S,Q},\nonumber\\[10pt]
&&H_{13}=H_{31}=\left(\frac{d^2M}{dSdQ}\right)_{P}=\left(\frac{dT}{dQ}\right)_{S,P}=\left[\left(\frac{\partial T}{\partial r_{_+}}\right)_{S,P}\left(\frac{dr_{_+}}{db}\right)_{S,P}+\left(\frac{\partial T}{\partial b}\right)_{S,P}\right]\left(\frac{dQ}{db}\right)^{-1}_{S,P}\nonumber\\[10pt]
&&H_{23}=H_{32}=\left(\frac{d^2M}{dQdP}\right)_{S}=\left(\frac{dM_Q}{dP}\right)_{S,Q}\nonumber\\&&\indent\indent\indent\,\,\,=\left(\frac{\partial M_Q}{\partial P}\right)_{S,Q}+\left[\left(\frac{\partial M_Q}{\partial r_{_+}}\right)_{S,Q}\left(\frac{dr_{_+}}{db}\right)_{S,Q}+\left(\frac{\partial M_Q}{\partial b}\right)_{S,Q}\right]\left(\frac{dP}{db}\right)^{-1}_{S,Q},\nonumber\\[10pt]
&&
\end{eqnarray}
together with the following definitions
\begin{eqnarray}
&&M_Q=\left(\frac{dM}{dQ}\right)_{S,P}=\left(\frac{\partial M}{\partial Q}\right)_{S,P}
+\left(\frac{\partial M}{\partial b}\right)_{S,P}\left(\frac{dQ}{db}\right)^{-1}_{S,P},\nonumber\\[10pt]
&&M_P=\left(\frac{dM}{dP}\right)_{S,Q}=\left(\frac{\partial
M}{\partial P}\right)_{S,Q}+\left(\frac{\partial M}{\partial
b}\right)_{S,Q}\left(\frac{dP}{db}\right)^{-1}_{S,Q}.
\end{eqnarray}
Using the above definitions, we can calculate the determinant of
Eq. (\ref{hessMat}). The final result is very complex, so we use
diagrams to analyze the stability. In all figures we set
$\Lambda=-3$.
\begin{figure}[H]
\centering\includegraphics[scale=.4]{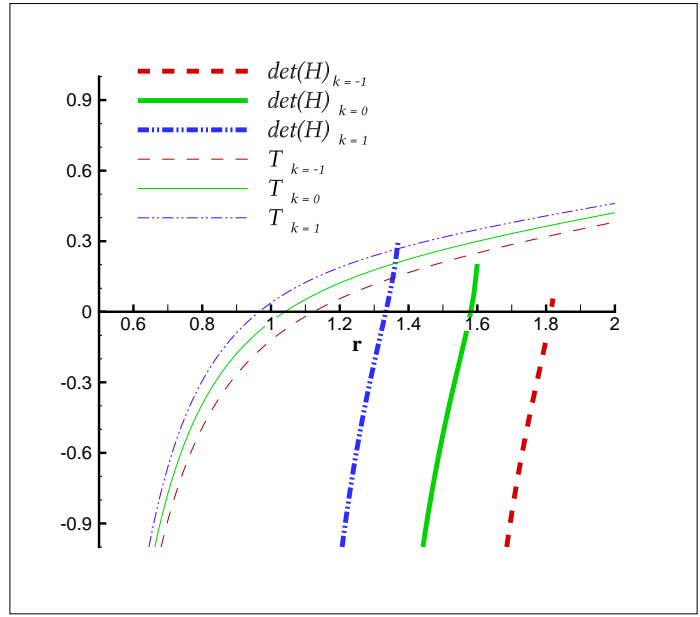}\caption{$b=0.2$, $P=1$ and $Q=0.6$.}\label{fig5}
\end{figure}
In Fig. \ref{fig5}, we plot the determinate for all three values
of topological constant $k$ respect to $r$. Since a real black
hole should have a non-zero value for temperature, so we should
have a careful look at the temperature in all considerations. In
this figure, the three bold lines refer to the determinate of
Hessian matrix which is displayed by \emph{det(H)} \footnote{we
use scale $10^{-1}$ for determinate function to better showing},
and the three thin lines relate to the temperature. According to
this figure, when we fix the metric parameters, all three
temperature functions enjoy a minimum at radius $r_{min_{_T}}$ for
which $r<r_{min_{_T}}$, this function is negative. Also, one may
see a minimum radius for the determinate functions $r_{min_d}$
that it is negative in the interval $r<r_{min_d}$. When
$r_{min_{_T}}<r<r_{min_d}$, the black hole is not thermally
stable. This range increases as the values of $k$ decreases. Our
calculations show that another special radius exists for the
determinate function that we call it ($r_{max_d}$) which is the
final point in which this function is displayed. We see that when
$r>r_{max_d}$ the values of determinate functions are imaginary so
we can not discuss in this interval using the Hessian matrix. As a
result, for a special values  of the metric parameters, the black
hole experiences a stable state if $r_{min_d}<r<r_{max_d}$ and
this interval changes as metric parameters change, for example by
increasing $k$ this stable range increases, too.
\begin{figure}[H]
    \centering \subfigure[$\,r=0.6$]{\includegraphics[scale=0.3]{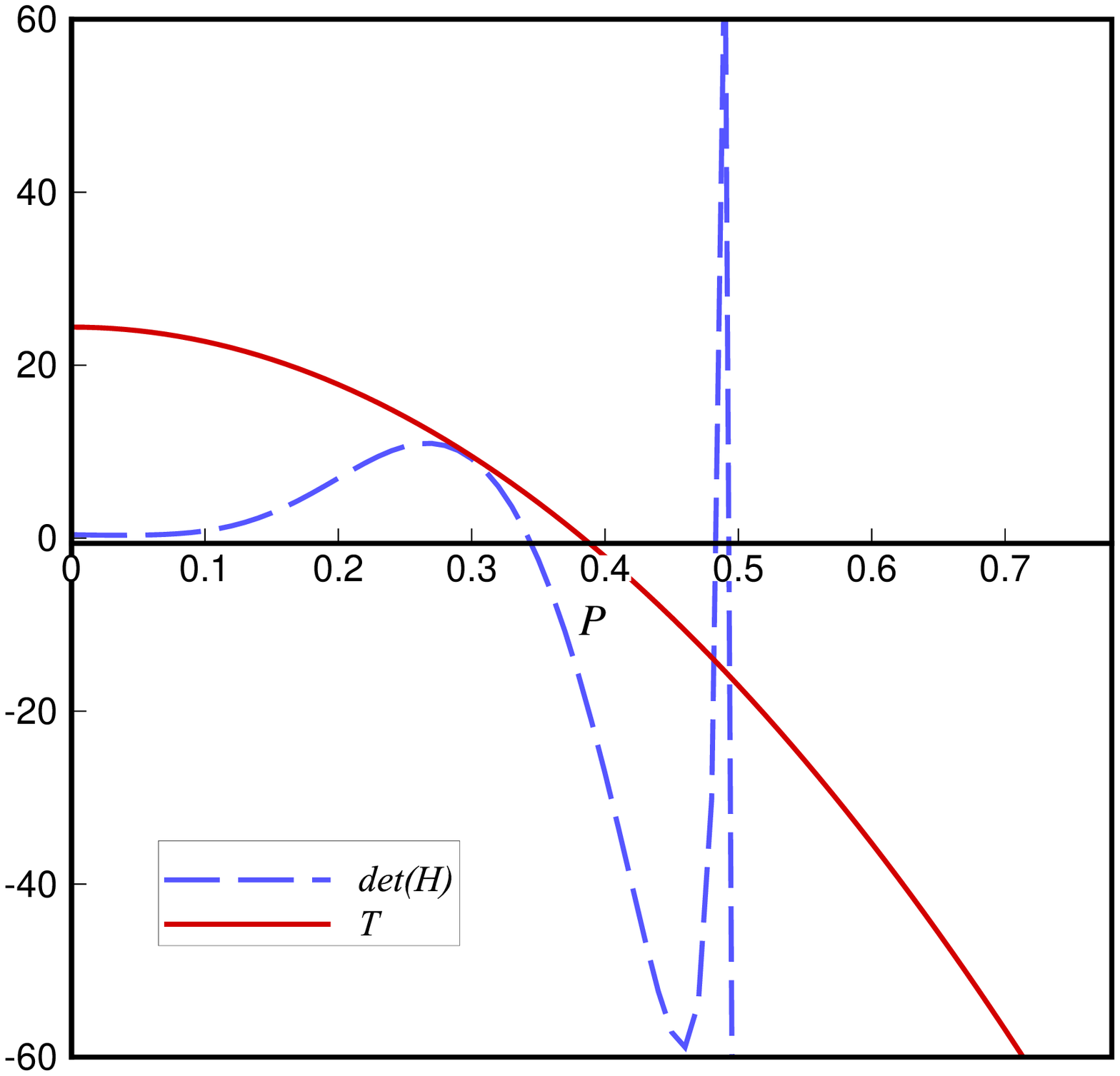}\label{fig6a}}
    \hspace*{.1cm} \subfigure[ $\,r=1$ ]{\includegraphics[scale=0.3]{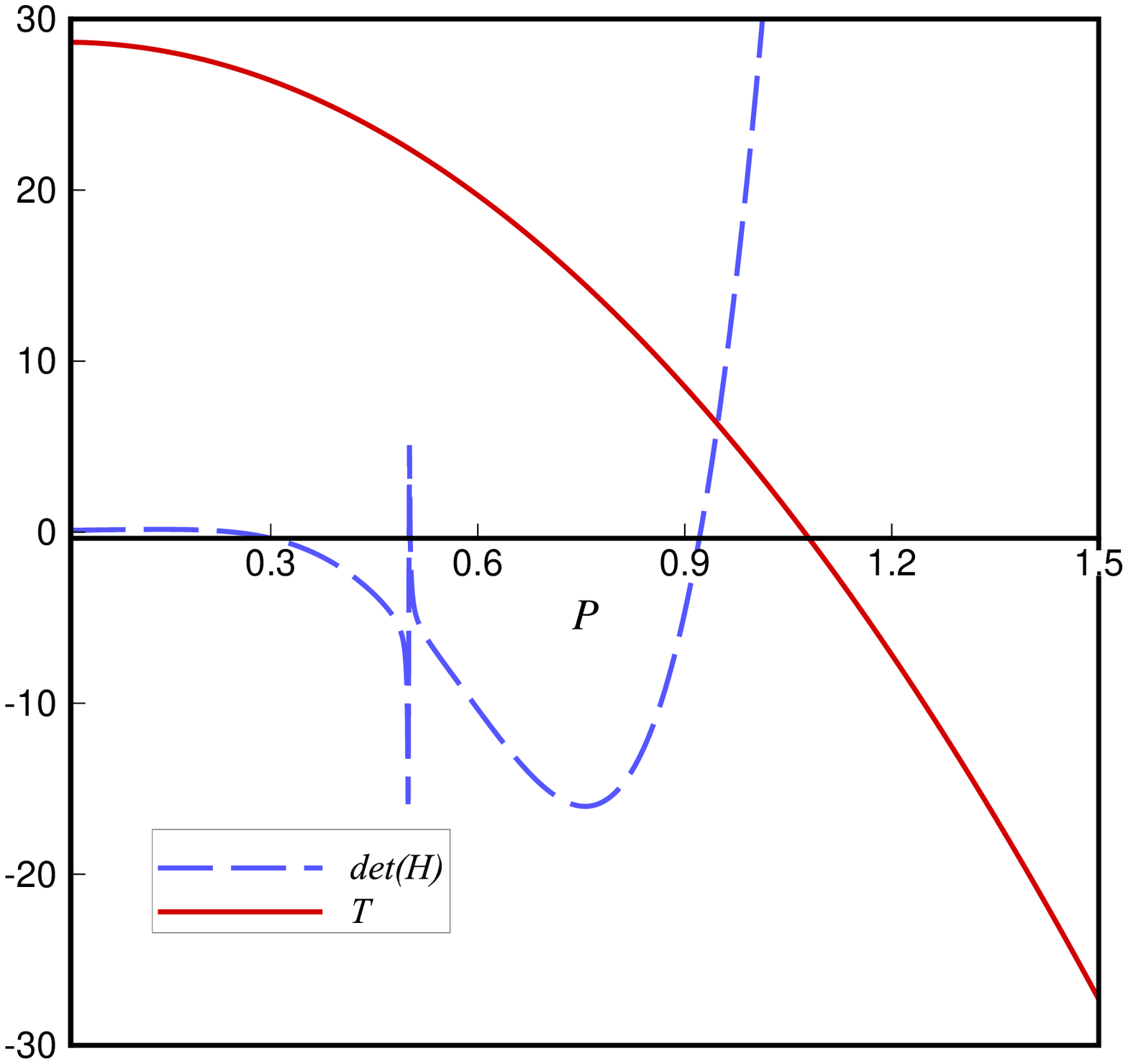}\label{fig6b}}
    \caption{$b=0.2$, $k=1$ and $Q=0.5$}\label{fig6}
\end{figure}
In order to better understanding the effects of magnetic charge on
the stability, we plot Fig. \ref{fig6}\footnote{We use scale
$10^{2}$ for temperature to better showing.}. In this diagram, we
fix the radius and allow the magnetic charge $P$ to change. By
fixing other metric parameters, one can see that in order to have
a real black hole which has positive values of temperature, there
exist a maximum for the magnetic charge ($P_{max_{_T}}$). Also, a
divergency in \emph{det(H)} is observed which corresponds to the
equality of charges ($P=Q$). As mentioned before, the positivity
of both temperature and Hessian determinate is necessary for
stability. Based on the figures, when $P<P_{max_{_T}}$ the
temperature is positive. The curve of \emph{det(H)} has some
special point ($P^i_d$) in this interval which shows that the
curve goes to zero at this point and its sign will change. The
numbers of this points increase as we consider the larger radius
for black hole. For instance, when $r=0.6$ one special point exist
($P^1_d$) and the stable interval places at $0<P<P^1_d$. But for
$r=1$, \emph{det(H)} becomes zero at two points. We call them
$P^1_d$ and $P^2_d$. Therefore, thermal stability occurs for two
smaller intervals ($0<P<P^1_d$ and $P^2_d<P<P_{max_{_T}}$).

In order to determine thermal stability and possible phase
transition of the black holes, one may check the heat capacity in
fixed values of electric charge.  In this case, we work in the
canonical ensemble.
\begin{equation}\label{CQ}
C_Q=T\left(\frac{dS}{dT}\right)_Q=\frac{1}{A}\left[3\, \left( -
\left( 2\,b-3\,r \right) r \left( b-r \right) ^{2}\Lambda -3\,
\left( b-r \right) ^{2}k+6\,{P}^{2} \right) {r}^{2} \left( b-r
\right) ^{2} \left( {P}^{2}-{q}^{2} \right)\right],
\end{equation}
where one can calculate the denominator as
\begin{eqnarray}
A&&=\left[-3\,{r}^{2}{b}^{2} \left( b-r \right) ^{4}\Lambda-18\,r
\left( b-r \right) ^{2} \left( 2\,b-r \right) {P}^{2}-18\, \left(
b-r \right) ^{4}{q}^{2} \right] k-{r}^{3} \left( b-r \right) ^{4}
\left( 2\,b-r \right) {b}^{2}{\Lambda}^{2}\nonumber\\&&+ \left[
-6\,{r}^{2} \left( {b}^{2} +4\,br-3\,{r}^{2} \right)  \left( b-r
\right) ^{2}{P}^{2}-6\,r \left( b-r \right) ^{4} \left( 2\,b+3\,r
\right) {q}^{2} \right] \Lambda+36\, r \left( 2\,b-3\,r \right)
{P}^{4}\nonumber\\&&-36\, \left( b+3\,r \right)  \left( b-r
\right) {q}^{2}{P}^{2}.
\end{eqnarray}
As we know, only the positive values of the thermal capacity of a
thermodynamic object are acceptable and the negative values show
that the object is not real. Also, In black hole physics in the
presence of positive temperature, divergency of heat capacity
shows that the black hole is not stable and needs a
thermodynamical phase transition. We call this transition as type
one. Also, zero values of $C_Q$ is not acceptable and usually
occurs for extremal black holes (which they have zero value for
temperature). The black hole may experience another type of phase
transition at this point which we call it type two. If the values
of heat capacity change to positive ones, it means that the
extremal black hole becomes a black hole with two horizons.

\begin{figure}[H]
\centering \subfigure[$\,Q=0.2$]{\includegraphics[scale=0.3]{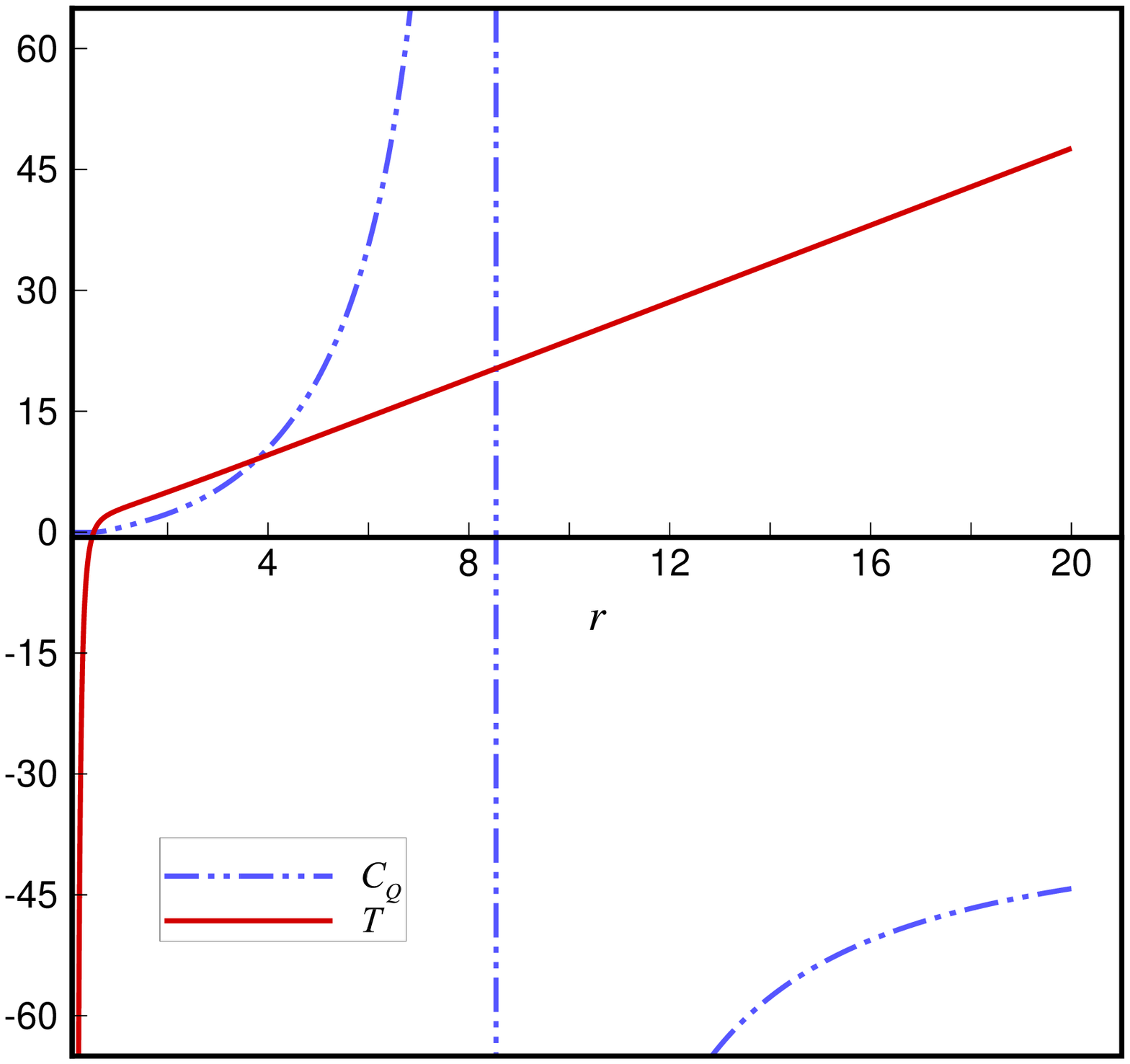}\label{fig7a}}
\hspace*{.1cm} \subfigure[ $\,Q=6$ ]{\includegraphics[scale=0.3]{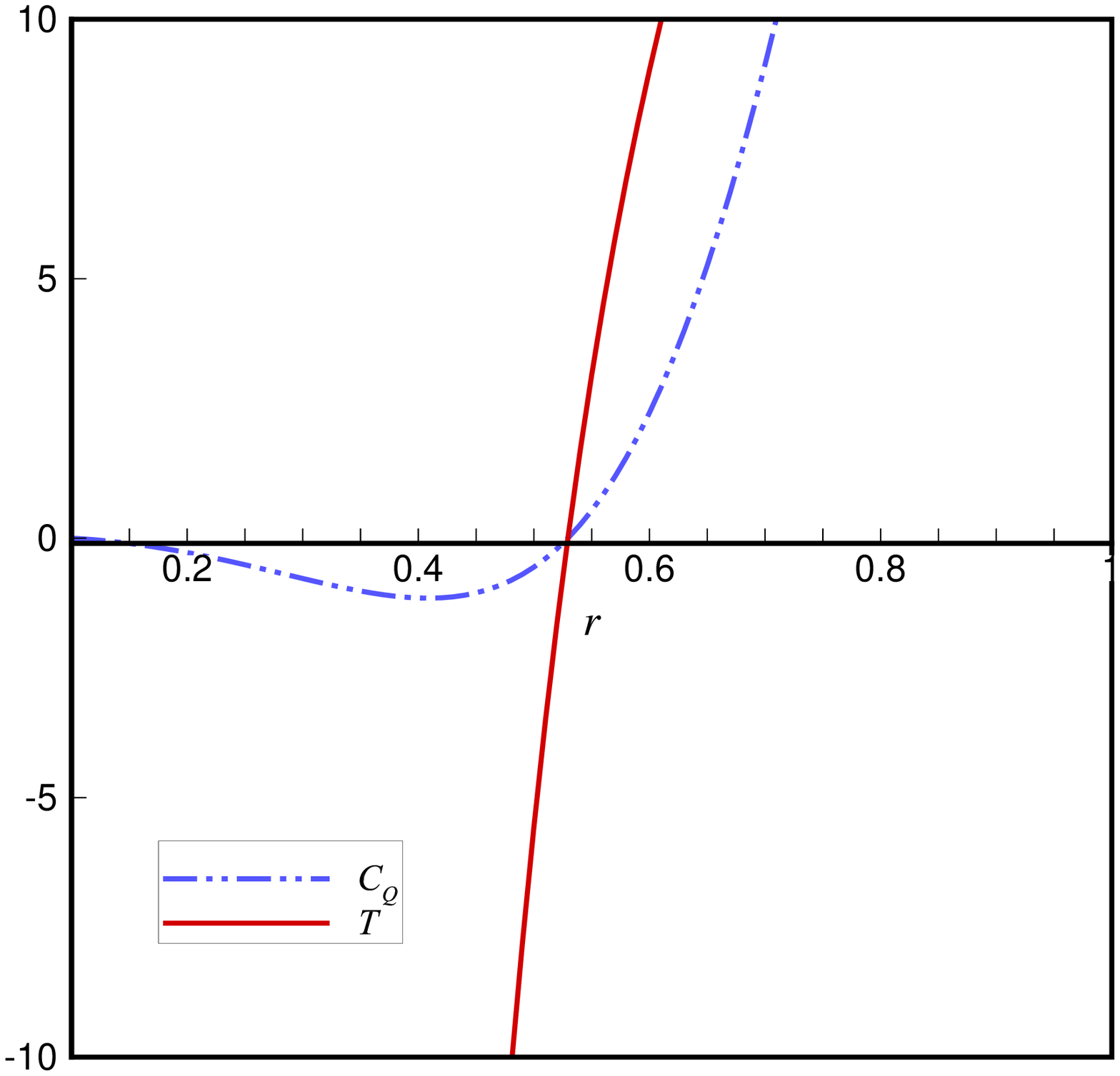}\label{fig7b}}
\caption{$b=0.1$, $k=1$ and $P=0.4$}\label{fig7}
\end{figure}

Looking at (\ref{CQ}), one may find that, regardless of other
metric parameters, by choosing equal values for both charges
($Q=P$), this function vanishes. We plot heat capacity and the
temperature in Fig.\ref{fig7} to better analyzing\footnote{We
multiplied both functions at 100 to better showing}. We consider
$\Lambda=-3$ in this figure.
By fixing other metric parameters, Fig.\ref{fig7a} shows that when
the magnetic charge is larger than electric charge ($P>Q$), both
types of phase transition occur. The divergency in $C_Q$ happens
at positive temperature interval. As the values of radius
increase, the heat capacity changes from positive to negative
values, and thermal phase transition makes the black hole smaller
(transition between small and large black hole). That's because,
for large $r$, heat capacity is negative and black hole cannot
exist in this range. Choosing smaller values for the magnetic
charge ($P<Q$), we only observe the type one of phase transition
(see Fig.\ref{fig7b}). Since the temperature values are also
becoming positive, we can deduce that after this point our black
hole will have two horizons. Note that for extremal black holes,
the temperature is zero. Finally, we conclude that black holes
stability are crucially dependent on the metric parameters.
\section{Closing remarks}\label{conclus}
In this paper, we have constructed a new class of topological
dyonic dilaton black hole solutions in the bachground of AdS
spacetime. The only case we could find an analytical solution
which fully satisfy all components of the field equations, is the
string case where the dilaton coupling constant is taken
$\alpha=1$. We have taken the dilaton potential in the form of the
combination of three Liouville type introduced in
\cite{gao1,gao2}. By solving the equation of motions, we obtained
the solutions which depend on both magnetic and electric charges.
We showed that the geometric mass of the black hole and two
charges are not independent of each other. Based on the metric
parameters, our solutions can be a black hole with two horizons,
extremal black hole or a naked singularity. Then, we explored
thermodynamics of these topological dilaton black holes and
checked the first law of thermodynamics. Finally, we disclosed the
effects of metric parameters on the stability of the solutions.
There are three extensive parameters (electric charge $Q$,
magnetic charge $P$ and entropy $S$) in our theory, it is
recommended to use Hessian matrix method to check the stability.
The positivity of temperature and the determinate of Hessian
matrix simultaneously, ensure that the solution is stable.

We found that the stability of our solutions is closely related to
the values of the metric parameters and it is not stable at all
interval of $r$. For example, when other metric parameters are
fixed, by increasing the value of the topological constant, $k$,
the stable interval decreases. Also, for the equal values of
electric and magnetic charges, the Hessian method did not work
since the determinate diverges.

We also checked the stability in the canonical ensemble, where the
electric charge is a fixed parameter. Thus, the positivity of the
thermal capacity $C_Q$ is sufficient to ensure the local
stability.  In our theory, we observed that, as other metric
parameters are fixed, when $Q<P$, the heat capacity diverges at a
special radius and type two of phase transition which coincides
with the divergence of $C_Q$ (while the temperature is positive)
occurs. It is important to note that $C_Q$ vanishes by considering
equal values for both electric and magnetic charges ($P=Q$) for
all radius. Also, we got the type two of phase transition
coincides with vanishing of the heat capacity and temperature, and
interpreted it as a transition from a non-physical black holes to
physical one or a transition from extremal black hole to black
hole with two horizons. By choosing $Q>P$ this transition is also
observed. As a result, all figures show that, the stability is
quite dependent on the metric parameters, and the change of them
can alter the black hole stability.
\acknowledgments{We thank Shiraz University Research Council. This
work has been supported financially by Research Institute for
Astronomy and Astrophysics of Maragha, Iran.}

\end{document}